\documentclass[twocolumn]{autart}    % Enable this line and disable the 
\usepackage{graphicx}          % Include this line if your 
\usepackage{amsmath}
\usepackage{amssymb}
\usepackage{helvet}
\usepackage[acronym, toc]{glossaries}
\usepackage{siunitx}
\usepackage{tikz,bm,color}
\usepackage{enumerate}
\usepackage{mathtools, cuted}
\usepackage{pgfplots}
\usepackage{pgfplotstable}
\usetikzlibrary{external,calc,patterns,decorations.pathmorphing,decorations.markings,decorations.text,arrows,shapes,positioning, backgrounds}
\usetikzlibrary{spy}
\usepgfplotslibrary{fillbetween}

\usepackage{hhline}
\usepackage{ctable}
\usepackage{pbox}

\definecolor{cadmiumgreen}{rgb}{0.0, 0.42, 0.24}
\definecolor{PIcolor}{rgb}{0, 0, 1.0}
\definecolor{STSMCcolor}{rgb}{0.929, 0.427, 0.067}
\definecolor{NACcolor}{rgb}{0.0, 0.42, 0.24}
\definecolor{InIcolor}{rgb}{0.247, 0.475, 0.749}
\definecolor{limitColor}{rgb}{0.247, 0, 0}
\pgfplotsset{compat=newest}
\usepackage{balance}

\newtheorem{assumption}{Assumption}

\newtheorem{remark}{Remark}
\newtheorem{proposition}{Proposition}

\renewcommand\qed{$\blacksquare$}
\usepackage{algorithm}
\usepackage{algpseudocode}
\usepackage{slashbox}
\newcommand\T{\rule{0pt}{2.6ex}}       % Top strut
\newcommand\B{\rule[-1.2ex]{0pt}{0pt}} % Bottom strut

\newacronym{FTC}{FTC}{Fault-Tolerant Control}
\newacronym{AC}{AC}{Alternating Current}
\newacronym{PMSM}{PMSM}{Permanent Magnet Synchronous Motor}
\newacronym{PMSMs}{PMSMs}{Permanent Magnet Synchronous Motors}
\newacronym{CNC}{CNC}{Computer Numerical Control}
\newacronym{SMO}{SMO}{Sliding Mode Observer}
\newacronym{SMC}{SMC}{Sliding Mode Control}
\newacronym{HOSMC}{HOSMC}{High-order Sliding Mode Control}
\newacronym{ISS}{ISS}{Input-to-State Stable}
\newacronym{LMI}{LMI}{Linear Matrix Inequality}
\newacronym{EKF}{EKF}{Extended Kalman Filter}
\newacronym{SISO}{SISO}{Single-Input Single-Output}
\newacronym{PID}{PID}{Proportional-Integral-Differential}
\newacronym{PI}{PI}{Proportional-Integral}
\newacronym{P}{P}{Proportional}
\newacronym{rpm}{rpm}{rounds pre minute}
\newacronym{STSMO}{STSMO}{Super-twisting Sliding Mode observer}
\newacronym{STSMC}{STS\-MC}{Super-twisting Sliding Mode controller}
\newacronym{PE}{PE}{Persistence of Excitation}
\newacronym{AO}{AO}{Adaptive Observer}
\newacronym{UGAS}{UGAS}{Uniformly Globally Asymptotically Stable}
\newacronym{BIBO}{BIBO}{Bounded Input-Bounded Output}
\newacronym{ULES}{ULES}{Uniformly Locally Exponentially Stable}
\newacronym{UB}{UB}{Uniformly Bounded}
\newacronym{UGB}{UGB}{Uniformly Globally Bounded}
\newacronym{MAEE}{MAEE}{Maximum Absolute Estimation Error}
\newacronym{VSC}{VSC}{Variable-Structure Control}
\newacronym{DC}{DC}{Direct Current}

\makenoidxglossaries

\begin{document}

\begin{frontmatter}

\title{On the Behaviour of Under-tuned Super-Twisting Sliding Mode Control Loops} % Title, preferably not more 

\thanks[footnoteinfo]{This paper was not presented at any IFAC 
meeting. Corresponding author D.~Papageorgiou. Tel. +45 45253573.}

\author[dimpa]{Dimitrios Papageorgiou}\ead{dimpa@elektro.dtu.dk},    % Add the 
\author[cedwards]{Christopher Edwards}\ead{c.edwards@exeter.ac.uk},  % e-mail address 

\address[dimpa]{Department of Electrical Engineering, Technical University of Denmark, Elektrovej 326, 2800 Kgs Lyngby, Denmark}  % Please supply                                              
\address[cedwards]{Department of Engineering, University of Exeter, Harrison Building, EX4 4QF Exeter, United Kingdom}             % full addresses here.

\begin{keyword}                           % Five to ten keywords,  
	Super-twisting sliding mode control; averaging; discontinuous dynamics; periodic perturbations; tuning               				  % chosen from the IFAC 
\end{keyword}                             % keyword list or with the 
                                          % help of the Automatica 
                                          % keyword wizard

\begin{abstract}                          % Abstract of not more than 200 words.
	This paper investigates the stability properties and performance of super-twisting sliding-mode control loops subject to periodic perturbations. Although there exist conditions on the control gains that guarantee finite-time stability of the closed-loop system, such conditions are often too restrictive from a practical standpoint, especially in relation to actuator limitations and induced chatter. Using regularisation and averaging theory, it is proven that under milder conditions for the control gains, the trajectories of the periodically perturbed closed-loop system converge to a stable limit cycle of the same period containing the origin. Additionally, guidelines for selecting the controller gains are provided based on bounds of the closed-loop system states. Finally, the theoretical findings are validated through simulations.
\end{abstract}

\end{frontmatter}

\section{Introduction}
	Over the last decades \gls{VSC} \cite{emelyanov1967variable} and, specifically, \gls{SMC} algorithms \cite{utkin1992a} have become very popular in a large area of applications, ranging from flight control and aviation systems to machine tools and robots. Their key features, i.e. robust rejection of bounded unknown disturbances and finite-time convergence of the control error to the origin, make sliding mode principles very attractive for control and estimation, as well as for fault-diagnosis \cite{Edwards2000} and fault-tolerant control \cite{Alwi2010}. Higher-order \gls{SMC} methods have also been researched in connection to alleviation of chattering effects \cite{bartolini661074}, which are present in conventional sliding modes. More specifically, the \gls{STSMC} introduced in \cite{levant1993sliding} and further detailed in \cite{Levant2002} and generalised in \cite{haimovich2019a}, has been shown to guarantee robust finite-time stabilisation and reduced chattering for appropriate selection of gains \cite{Levant2007576,Moreno2012}.

Systematic tuning of the \gls{STSMC} based on performance specifications is highly desirable, especially in industrial applications since it facilitates easy commissioning of control systems and guarantees that certain levels of accuracy are achieved. Although the design complexity of the \gls{STSMC} is relatively low, the selection of the control gains has received significant attention. Strict Lyapunov functions were designed in \cite{Moreno2012} for ascertaining finite-time convergence of the \gls{STSMC} closed loop to the origin. Explicit expressions for the controller gains were provided, based on which, an estimation of the reaching time was obtained. Geometric arguments proving finite-time stability of the \gls{STSMC} were provided in \cite{behera2018new}, where the controller gains needed to satisfy certain bounding conditions. A similar approach was followed in \cite{seeber2018necessary}, in which the authors provided necessary and sufficient conditions for finite-time convergence to the origin. The \gls{STSMC} loop was tuned based on the requirement that the majorant curve had to contract towards the origin. Describing functions were used in \cite{pilloni2012a} for the selection of the \gls{STSMC} gains and the tuning rules were based on specifications for the properties of limit cycles that appear in linear systems with unmodelled actuator dynamics. Additional studies that addressed the issue of conservative controller gains focused on variable-gain variations of the \gls{STSMC} algorithm. The application of such an adaptive \gls{STSMC} design to an electropneumatic actuator was presented in  \cite{shtessel2012a}. Finite-time convergence was proven and the reaching time was calculated. A more recent approach to estimate the reaching time was given in \cite{seeber2018a}. The combination of a certainty equivalence controller with an adaptive gain \gls{STSMC} was pursued in \cite{barth2015a}, where the scheme was used to avoid unnecessary large controller gains. A dual-layer adaptive \gls{STSMC} was presented in \cite{edwards2016adaptive} and \cite{edwards2016a}. Adaptation laws that guarantee finite-time convergence to the origin were provided for both known and unknown perturbation bounds.

Most of the aforementioned studies provide tuning rules for the \gls{STSMC} that guarantee finite-time convergence. However, they end up proposing gains that can be significantly large. \textit{A requirement that is common to all these works pertains to the integral gain of the \gls{STSMC} being larger than the perturbation rate bound.} This can be very limiting in terms of actuator operation ranges and chattering levels, especially in electromechanical systems. To understand this, consider a simplified model of the angular velocity dynamics of a drive \gls{PMSM} in a linear axis system given by
 %%%%%%%%%%%%%%%%%%%%%%%%%%%%%
\begin{align*}
	\dot{\omega} &= \frac{1}{J}\left[ u - \underbrace{T_C\frac{2}{\pi}\arctan(\alpha \omega) - \beta\omega}_{T_F(\omega)} - T_L \right], \; \alpha > 100
\end{align*}
where $T_F$ is the uncertain Coulomb and viscous friction torque and $T_L$ is the known load torque \cite{papageorgiou2019adaptive}. The closed-loop dynamics of the tracking error $e(t) \triangleq \omega(t) - r(t)$, where $r(t)$ is a sufficiently smooth demand signal using the \gls{STSMC} law $u(t) = T_L + J\dot{r}(t) - k_1\vert e(t) \vert^{\frac{1}{2}}\text{sgn}(e(t)) - k_2\int_0^{t}\text{sgn}(e(\tau))d\tau$, is given by
\begin{equation}
	\dot{e} = \frac{1}{J}\left[ - k_1\vert e \vert^{\frac{1}{2}}\text{sgn}(e) - k_2\int_0^{t}\text{sgn}(e(\tau))d\tau - T_F \right]
\end{equation}
Ensuring finite-time convergence of the tracking error to the origin requires that $k_2 > \left \vert \dot{T}_F \right\vert$ \cite{Moreno2012,seeber2018necessary}, with
\begin{equation}\label{eq:friction_derivative}
	\dot{T}_F = \frac{\partial T_F}{\partial \omega}\dot{\omega} = \left[ \beta + \frac{2\alpha T_C}{\pi\left( 1 + \alpha^2\omega^2 \right)} \right]\frac{u - T_F(\omega) - T_L}{J}
\end{equation}
Equation \eqref{eq:friction_derivative} shows that apart from then inherent difficulty in calculating a not overly conservative bound for $\dot{T}_F$ due to the algebraic loop (the gains for the \gls{STSMC} in $u$ depend on $u$ itself), $k_2$ must assume (unrealistically) large values since $\dot{T}_F$ becomes very big for velocities close to zero (because of the steepness factor $\alpha$). However, despite these limitations it has been experimentally shown \cite{papageorgiou2019adaptive} that under an appropriate selection of the gains, the \gls{STSMC} can outperform conventional and several advanced control schemes in industrial applications, even if $k_2 < \left \vert \dot{T}_F \right\vert$.
%%%%%%%%%%%%%%%%%%%%%%%%%%%%%%%%%%%%%%%%%%%%%%%%

The tuning challenge is related to the fact that the theoretical condition for the selection of $k_2$ stems from the requirement of finite-time convergence of the controlled variable to the origin, which is very demanding for real physical systems. In fact, \textit{ensuring boundedness with prescribed accuracy bounds} can be sufficient for practical applications that relate to tracking and positioning. Of particular interest are the cases where the system dynamics is affected by periodic perturbations. Such is the case of friction forces and cogging torques during motion reversals of mechanical components in industrial applications that include repeated closed-curve tracking \cite{altintas2001manufacturing}. For instance, the motion profile of a machine tool drive axis during a contouring task consists of periodic segments that include axis reversals. The effect of Coulomb friction and stiction on the drive motor and axis dynamics can be described as a periodic torque or force perturbation that cause contouring deformations \cite{gross2001electrical}. This is the engineering motivation for the work in this paper.

The main contribution of this paper is that investigates the stability properties of \gls{SISO} \gls{STSMC} closed-loop systems with bounded-rate periodic perturbations, where the conditions that guarantee finite-time convergence no longer hold. Specifically:
\begin{itemize}
	\item It is shown that under milder (smaller) gain conditions, the solutions of the closed-loop system converge to a stable limit cycle around the origin, which has the same period as the perturbation.
	\item Bounds for the control error are derived as functions of the controller gains and the perturbation characteristics.
	\item Guidelines for the tuning  of the \gls{STSMC} are provided and validated in simulations.
\end{itemize} 

The remainder of the paper is structured as follows: Section \ref{sec:preliminaries} introduces the class of perturbed systems that are examined in this study and states the underlying assumptions. The analysis of the closed-loop system  is carried out in Section \ref{sec:analysis}, where the stability properties of the system are described and proven. Section \ref{sec:tuningRules} provides tuning guidelines on the selection of the \gls{STSMC} gains based on derived bounds on the controlled variable. Simulation results that verify the theoretical findings are presented in Section \ref{sec:simulations} and finally, concluding remarks are given in Section \ref{sec:conclusions}.

\section{Preliminaries} \label{sec:preliminaries}
	This study considers the scalar nonlinear systems described by the differential equation
\begin{equation}
	\dot{y} = h(t,y) + g(t,y)u_0 + d(t)
\end{equation}
where $y\in\mathbb{R}$ is available from measurements, the scalar functions $h(t,y),g(t,y)\in\mathcal{C}^1$ are bounded for bounded $y$, $g(t,y) \neq 0, \; \forall (t,y)\in[0,\infty)\times\mathbb{R}$ and $d(t)\in\mathcal{C}^2$ is a $T$-periodic perturbation. Selecting a control law
\begin{align}
	u_0 &= g^{-1}(t,y)\left[ -h(t,y) + u \right]\\
	u &= -k_1\vert y \vert^{\frac{1}{2}}\text{sgn}(y) - k_2\int_0^t \text{sgn}(y(\tau))d\tau \label{eq:control_law}
\end{align}
where \textnormal{sgn}$(\cdot)$ represents the signum function, results in the following closed-loop dynamics
\begin{equation}\label{eq:closed_loop_system}
	\boldsymbol{\dot{x}} = \boldsymbol{f}(t,\boldsymbol{x}) \; , \text{with}
\end{equation}
\begin{align}\label{eq:discontinuous_vector_field}
	\boldsymbol{f}(\boldsymbol{x}) &\triangleq \begin{bmatrix}
		-k_1\vert x_1 \vert^{\frac{1}{2}}\text{sgn}(x_1) + x_2\\
		-k_2\text{sgn}(x_1) + q(t)
	\end{bmatrix},\\
	\boldsymbol{x} &= \begin{bmatrix}
		x_1\\
		x_2
	\end{bmatrix} \triangleq \begin{bmatrix}
		y\\
		-k_2\int_0^t \text{sgn}(y(\tau))d\tau + d(t)
	\end{bmatrix} \nonumber
\end{align}
and $q(t) \triangleq \dot{d}(t)$. Since $d(t)\in\mathcal{C}^2$ and is $T$-periodic, it follows that its derivative is a continuous bounded $T$-periodic function. Let $\vert q(t) \vert \leq L$, where $L > 0$. It has been shown \cite{seeber2018necessary} that if
\begin{align}
	k_2 &> L \label{eq:finite_time_conditions_k_2}\\
	k_1 &\geq 1.8\sqrt{k_2 + L}, \label{eq:finite_time_conditions_k_1}
\end{align}
then the system in \eqref{eq:closed_loop_system} has a unique finite-time stable equilibrium point at the origin, irrespectively of the perturbation type. The same holds for any $k_1,k_2 > 0$ in the case where $q(t) \equiv 0$ \cite{Moreno2012}. When $k_2 < L$, not only can finite-time stability not be guaranteed, but also the solutions of the closed-loop system may grow unbounded. Take for example the case where $q(t) = L > k_2$. Then the dynamics of $x_2$ reads: $\dot{x}_2 = L - k_2\text{sgn}(x_2) \geq L - k_2 > 0$, which implies that $\lim\limits_{t \rightarrow \infty} x_2(t) = \infty$. However, there can be cases where the solutions of the closed-loop system \eqref{eq:closed_loop_system} are bounded even though $k_2 < L$. This is very important in practical applications of the \gls{STSMC} since it means that acceptable performance may be achievable under less demanding gain requirements. As it will be shown in the next section, such cases include systems where the perturbation (and its derivative) are continuous $T$-periodic functions. Specifically, the conditions for the controller gains under which the solutions of a periodically perturbed \gls{STSMC} closed-loop system remain bounded will be studied along with the stability properties of such a system. The following assumption is adopted for the rest of the study:

\begin{assumption} \label{ass:contorller_gains}\normalfont
	The \gls{STSMC} gains $k_1,k_2$ satisfy $k_1 > 0$ and $0 < k_2 < L$.
\end{assumption}

\section{Closed-loop system analysis} \label{sec:analysis}
	The stability properties of the periodically perturbed closed-loop system will be analysed in this section. The main idea is to employ tools from averaging theory. Contrary to existing developments in the field of averaging for systems with discontinuous vector fields \cite{llibre2015a,llibre2015birth,llibre2012averaging} that rely on geometrical arguments, the approach used in this paper pertains to working with a continuous version of the original vector field, i.e. a \emph{regularisation} of it \cite{utkin1992a}. This allows the use of more straightforward tools from averaging theory for continuous vector fields. The implication of this choice is that the regularised system constitutes an approximation of the original system dynamics given in Equation \eqref{eq:closed_loop_system}. Showing that the regularisation provides a uniform approximation of all possible behaviours of the original system is not a trivial task and it is outside the scope of the paper. However, it does capture all possible behaviours of the \gls{STSMC} closed loops that emerge in real applications, where at best the signum function is approximated by a continuous function. The next Proposition shows that such an approximation can be made with arbitrarily high accuracy.
	\begin{proposition}\label{prop:regulization}
		Consider the system defined in Equation \eqref{eq:closed_loop_system}, where the gains $k_1,k_2$ are selected according to \eqref{eq:finite_time_conditions_k_2},\eqref{eq:finite_time_conditions_k_1}. There exists a regularisation $\boldsymbol{f}_{\delta}(t,\boldsymbol{x})$ of the discontinuous vector field $\boldsymbol{f}(t,\boldsymbol{x})$ defined in \eqref{eq:discontinuous_vector_field} such that for any finite $\delta > 0$ any solution $\boldsymbol{x}_{\delta}(t;t_0,\boldsymbol{x}_{\delta,0})$ of $\boldsymbol{\dot{x}}_{\delta} = \boldsymbol{f}_{\delta}(t,\boldsymbol{x}_{\delta})$ starting at $\boldsymbol{x}_{\delta,0} = \boldsymbol{x}_{\delta}(t_0)$ converges in finite time $t_f$ to a region containing the origin with area depending on $\delta$. The solution is contained in this region $\forall t > t_f$ and $\lim\limits_{\delta \rightarrow 0}\boldsymbol{x}_{\delta}(t;t_0,\boldsymbol{x}_{\delta,0}) = \boldsymbol{x}(t;t_0,\boldsymbol{x_0}) = \boldsymbol{0}$ with $\boldsymbol{x}(t;t_0,\boldsymbol{x_0})$ being the solution to \eqref{eq:closed_loop_system} starting at $\boldsymbol{x_0} = \boldsymbol{x}(t_0)$.
	\end{proposition}
	\begin{pf}
		Let  $\boldsymbol{f}_{\delta}(t,\boldsymbol{x})$ be a regularisation of  $\boldsymbol{f}(t,\boldsymbol{x})$ defined as
		\begin{align}\label{eq:regularised_vector_field}
			\boldsymbol{f}_{\delta}(t,\boldsymbol{x}) &\triangleq \begin{bmatrix}
				-k_1\vert x_1 \vert^{\frac{1}{2}}\phi_{\delta}(x_1,\delta) + x_2\\
				-k_2\phi_{\delta}(x_1,\delta) + q(t)
			\end{bmatrix}
		\end{align}
		where $\delta > 0$ and the continuous function $\phi_{\delta}:\mathbb{R}\times (0,+\infty) \rightarrow [-1,1]$ is defined as \cite{llibre2015birth}
		\begin{equation}
			\phi_{\delta}(q,\delta) \triangleq \begin{cases}
				1 &\text{ if } q \geq \delta\\
				\displaystyle\frac{q}{\delta} &\text{ if } -\delta < q < \delta\\
				-1 &\text{ if } q \leq -\delta
			\end{cases} \; .
		\end{equation} 
		The regularised system can be written as
		\begin{align}
			\dot{x}_1 &= -k_1\vert x_1 \vert^{\frac{1}{2}}\text{sgn}(x_1) + x_2 + k_1\vert x_1 \vert^{\frac{1}{2}} \rho(x_1,\delta) \label{eq:regularised_1}\\
			\dot{x}_2 &= -k_2\text{sgn}(x_1) + q(t) + k_2\rho(x_1,\delta) \label{eq:regularised_2}
		\end{align}
		where $\rho(x_1,\delta) \triangleq \text{sgn}(x_1) - \phi_{\delta}(x_1,\delta)$. It is easy to show that the following properties for $\rho$ hold:
		\begin{enumerate}
			\item $\vert \rho(x_1,\delta)  \vert \leq 1, \; \forall x_1\in \mathbb{R}, \; \forall \delta > 0$
			\item $\rho(x_1,\delta)  = 0, \; \forall x_1\in ( -\infty,-\delta]\cup[\delta,+\infty)$
			\item $\lim\limits_{\delta \rightarrow 0}\rho(x_1,\delta) = 0, \; \forall x_1\in\mathbb{R} - \{0\}$
		\end{enumerate}
		As in \cite{Moreno2012}, introducing the transformation
		$$
			\boldsymbol{z} = \begin{bmatrix}
				z_1\\
				z_2
			\end{bmatrix} \triangleq \begin{bmatrix}
				\vert x_1 \vert^{\frac{1}{2}}\text{sgn}(x_1)\\
				x_2
			\end{bmatrix}
		$$
		allows for re-writing the regularised system as
		\begin{equation} \label{eq:regularised_transformed}
			\boldsymbol{\dot{z}} = \vert x_1 \vert^{-\frac{1}{2}}\boldsymbol{A}\boldsymbol{w} + \boldsymbol{q}(t) + \boldsymbol{\xi}(x_1,\delta)
		\end{equation}
		when $x_1 \neq 0$ with
		$$
			\boldsymbol{A} = \begin{bmatrix}
				-\frac{1}{2}k_1 & \frac{1}{2}\\
				-k_2 & 0
			\end{bmatrix}, \; \boldsymbol{q}(t) = \begin{bmatrix}
				0\\
				q(t)
			\end{bmatrix}, \; \boldsymbol{\xi}(x_1,\delta) = \begin{bmatrix}
				\frac{1}{2}k_1\\
				k_2
			\end{bmatrix}\rho(x_1,\delta)
		$$
		When $\rho(x_1,\delta) \equiv 0$, i.e. $\delta = 0$, it has been shown \cite{Moreno2012} that provided that $q(t)$ is globally bounded, there exists selection of the gains $k_1,k_2$, that ensure the existence of positive definite matrices $$
			\boldsymbol{P} \triangleq \begin{bmatrix}
				p_1 & p_2\\
				p_2 & p_3
			\end{bmatrix}, \; \boldsymbol{Q} \triangleq \begin{bmatrix}
				q_1 & q_2\\
				q_2 & q_3
			\end{bmatrix}
		$$
		with $p_1,p_3,q_1,q_3 > 0$ and an absolutely continuous Lyapunov function $V_0 = \boldsymbol{z}^T\boldsymbol{P}\boldsymbol{z}$, with the property that
		$\dot{V}_0 \leq -\vert z_1 \vert^{-1} \boldsymbol{z}^T\boldsymbol{Q}\boldsymbol{z} \leq -c\sqrt{V_0}, \; c > 0$. Consider $V = \boldsymbol{z}^T\boldsymbol{P}\boldsymbol{z}$ as a Lyapunov function candidate for the regularised system in Equation \eqref{eq:regularised_transformed}. Its time derivative along the trajectories of \eqref{eq:regularised_transformed} is given by:
		\begin{align}
			\dot{V} &= \boldsymbol{\dot{z}}^T\boldsymbol{P}\boldsymbol{z} + \boldsymbol{z}^T\boldsymbol{P}\boldsymbol{\dot{z}} = \dot{V}_0 + 2\boldsymbol{z}^T\boldsymbol{P}\boldsymbol{\xi} \nonumber\\
			&\leq -\vert z_1 \vert^{-1} \boldsymbol{z}^T\boldsymbol{Q}\boldsymbol{z} + 2 \boldsymbol{z}^T\boldsymbol{P}\boldsymbol{\xi} \text{ , for } z_1 \neq 0 \; . \nonumber
		\end{align}
		Two situations will now be considered: If $\vert x_1 \vert \geq \delta$, then $\boldsymbol{\xi}(x_1,\delta) = \boldsymbol{0}$ and $\dot{V} \leq -c\sqrt{V} < 0$ \cite{Moreno2012}. In the situation when $\vert x_1 \vert < \delta$ or equivalently $\vert z_1 \vert < \sqrt{\delta}$ then
		\begin{equation}
			\dot{V} \leq -\frac{1}{\sqrt{\delta}}\boldsymbol{z}^T\boldsymbol{Q}\boldsymbol{z} + 2 \boldsymbol{z}^T\boldsymbol{P}\boldsymbol{\xi} \; .
		\end{equation}
		In order to find conditions for $\dot{V} < 0$  (when $\vert x_1 \vert < \delta$), it suffices to investigate where the inequality
		\begin{equation}\label{eq:lyapunov_conditions}
			-\boldsymbol{z}^T\boldsymbol{Q}\boldsymbol{z} + 2\sqrt{\delta} \boldsymbol{z}^T\boldsymbol{P}\boldsymbol{\xi} < 0
		\end{equation}
		holds. Notice that although $\boldsymbol{\xi}$ is a function of the state vector, it is globally bounded since
		$$
			\Vert \boldsymbol{\xi}(x_1,\delta) \Vert = \sqrt{\frac{1}{4}k_1^2 + k_2^2}\vert \rho(x_1,\delta) \vert \leq \sqrt{\frac{1}{4}k_1^2 + k_2^2} \triangleq \bar{\kappa}
		$$
		for a given $\delta > 0$ and $\forall x_1\in\mathbb{R}$. When $\sqrt{\vert x_1 \vert} < \sqrt{\delta}$, the left hand side of \eqref{eq:lyapunov_conditions} can be expanded as:
		\begin{align*}
			&-q_1 \vert x_1 \vert - 2\sqrt{\vert x_1 \vert}\text{sgn}(x_1) x_2q_2 - q_3 x_2^2\\
			&+ 2\sqrt{\delta} \sqrt{\vert x_1 \vert}\text{sgn}(x_1)\left( p_1 \xi_1 + p_2 \xi_2 \right) + 2\sqrt{\delta} x_2\left( p_2 \xi_1 + p_3 \xi_2 \right)\\
			&\leq -q_3 \vert x_2 \vert^2 - q_1\vert x_1 \vert + 2\sqrt{\vert x_1 \vert} \vert q_2 \vert \vert x_2 \vert\\
			&+ 2\sqrt{\delta} \sqrt{\vert x_1 \vert}\sqrt{p_1^2 + p_2^2}\Vert \boldsymbol{\xi} \Vert + 2\sqrt{\delta} \vert x_2 \vert \sqrt{p_2^2 + p_3^2}\Vert \boldsymbol{\xi} \Vert\\
			&\leq -q_3 \vert x_2 \vert^2 + \mu \vert x_2 \vert + \nu \triangleq \Lambda\left( \vert x_2 \vert \right)
		\end{align*}
		with the coefficients of the polynomial $\Lambda\left( \vert x_2 \vert \right)$ defined as
		\begin{align*}
			\mu &\triangleq 2\sqrt{\delta}\left( \vert q_2 \vert + \bar{\kappa} \sqrt{p_2^2 + p_3^2} \right), \;\nu \triangleq 2\delta \bar{\kappa}\sqrt{p_1^2 + p_2^2} \; .
		\end{align*}
		Since the discriminant of $\Lambda\left( \vert x_2 \vert \right)$ is positive for $\delta > 0$, there exist two distinct real roots $\varrho_{1}(\delta), \varrho_{2}(\delta)$ with
		\begin{align*}
			&\max(\vert \varrho_1 \vert, \vert \varrho_2 \vert) = \frac{\sqrt{\delta}}{q_3}\left\vert \vert q_2 \vert + \bar{\kappa} \sqrt{p_2^2 + p_3^2} \right.\\
			&\left. + \sqrt{\left( \vert q_2 \vert + \bar{\kappa} \sqrt{p_2^2 + p_3^2} \right)^2 + 2q_3\bar{\kappa}\sqrt{p_1^2 + p_2^2}} \right\vert \triangleq \bar{\rho}\sqrt{\delta}
		\end{align*}
		such that $\Lambda\left( \vert x_2 \vert \right) < 0, \; \forall \vert x_2 \vert > \max(\vert \varrho_1 \vert, \vert \varrho_2 \vert)$. This, in turn, implies that for every $\delta > 0$ there exists $0 < \varepsilon_0(\delta) \leq \max(\vert \varrho_1 \vert, \vert \varrho_2 \vert)$, such that for every $\vert x_2 \vert > \varepsilon_0$, $\Lambda(\vert x_2 \vert) < 0$ and hence, $\dot{V} < 0$. It follows that outside the rectangle $\mathcal{W} = \{ (x_1,x_2)\in\mathbb{R}^2 \big | \vert x_1 \vert \leq \delta, \; \vert x_2 \vert \leq \varepsilon_0 \}$ (see Figure \ref{fig:rectangle}) $\dot{V} < 0$ holds. Let $\mathcal{V}_r$ be the smallest ellipsoid of the form $\mathcal{V}_r = \{z | V(z)\leq r \}$ such that $\mathcal{W} \subset \mathcal{V}_r$. Then outside $\mathcal{V}_r$ the function $V$ is decreasing and so $\mathcal{V}_r$ is an invariant set. Furthermore, the area $\mathcal{A}$ of the rectangle satisfies $\mathcal{A} \leq 4\delta\sqrt{\delta}\bar{\rho}$, which is a continuous, strictly increasing function of $\delta$ and therefore  $\mathcal{W}$ and hence $\mathcal{V}_r$ can be made arbitrarily small as $\delta \rightarrow 0$. \qed
	\end{pf}
	\begin{figure}[t]
		\begin{center}
			\includegraphics[width = 0.5\textwidth]{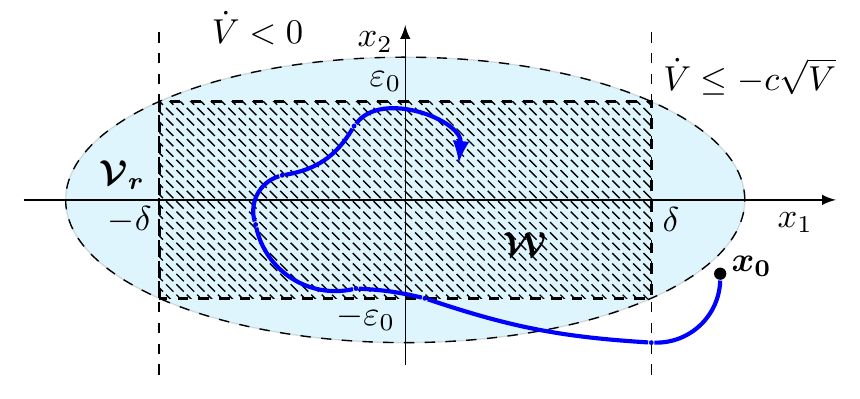}
			\vspace{-11pt}
			\caption{The trajectories are confined in a rectangle $\mathcal{W}$ with area equal to $\mathcal{A}(\delta) = 4\delta\varepsilon_0(\delta)$ that tends to 0 as $\delta \rightarrow 0$.}\label{fig:rectangle}
		\end{center}
	\end{figure}
	The main result of this paper pertains to showing the existence of a unique limit cycle in the dynamics of the approximated system and it is stated in the following proposition.
	\begin{proposition}
		Consider the closed-loop system \eqref{eq:closed_loop_system} and its approximation associated with the regularisation \eqref{eq:regularised_vector_field}, where $q(t)$ is Lipschitz, $T$-periodic of sufficiently small period $T$ and $\vert q(t) \vert \leq L$. Then, $\exists \varepsilon_1 > 0$ with  $0 < T <\varepsilon_1$ such that under the conditions 
		\begin{align}
			k_2 &> \left\vert \frac{1}{T}\int_{0}^{T}q(t)dt \right\vert \label{eq:convergence_condition_k2}\\
			k_1 &\geq 1.8\sqrt{k_2 + \left\vert \frac{1}{T}\int_{0}^{T}q(t)dt \right\vert} \label{eq:convergence_condition_k1}
		\end{align}
		the trajectories of the regularised system $\boldsymbol{\dot{x}} = \boldsymbol{f}_{\delta}(t,\boldsymbol{x})$ converge to a limit cycle with period $T$.
	\end{proposition}
	\begin{pf}
		The regularised system can be written as
		\begin{equation}\label{eq:averaging_form}
			\boldsymbol{\dot{x}} = \varepsilon \frac{1}{T}\boldsymbol{f}_{\delta}(t,\boldsymbol{x}) \triangleq \varepsilon \boldsymbol{g}(t,\boldsymbol{x}) \; , \; \varepsilon = T \; ,
		\end{equation}
		where $\boldsymbol{g}(t,\boldsymbol{x})$ is obviously Lipschitz continuous and $T$-periodic. The associated averaged system is written as
		\begin{equation}
				\boldsymbol{\dot{\chi}} = \varepsilon \boldsymbol{\bar{g}}(\boldsymbol{\chi}), \; \boldsymbol{\chi} = \begin{bmatrix}
				\chi_1 & \chi_2
				\end{bmatrix}^T\in \mathbb{R}^2
			\end{equation}
			with $\varepsilon \boldsymbol{\bar{g}}(\boldsymbol{\chi}) = \displaystyle\frac{1}{T}\int_{0}^{T}\boldsymbol{f}_{\delta}(t,\boldsymbol{x}) dt$ and finally
			\begin{equation}\label{eq:averaged_system}
				\boldsymbol{\dot{\chi}} = \begin{bmatrix}
				-k_1\vert \chi_1 \vert^{\frac{1}{2}}\phi_{\delta}(\chi_1,\delta) + \chi_2\\
				-k_2\phi_{\delta}(\chi_1,\delta) + \displaystyle\frac{1}{T}\int_{0}^{T}q(t) dt
				\end{bmatrix} \; .
			\end{equation}
		Comparing \eqref{eq:averaged_system} to \eqref{eq:regularised_vector_field} and taking Proposition \ref{prop:regulization} into consideration, reveals that if conditions \eqref{eq:convergence_condition_k2} and \eqref{eq:convergence_condition_k1} are satisfied, then for sufficiently small $\delta$ ($\delta \rightarrow 0$) the origin is a finite-time stable equilibrium point of the averaged system. Then, by Theorem 4.1.1 in \cite{guckenheimer1983a}, there exists $\varepsilon_1 > 0$, such that $\forall \varepsilon\in(0,\varepsilon_1)$, the solutions of \eqref{eq:averaging_form} converge to a unique isolated $T$-periodic orbit $\gamma_{\varepsilon}(t) = O(\varepsilon)$. \qed
	\end{pf}
	\begin{remark}\normalfont
		The constant $\varepsilon_1$ can be associated with the largest time scale $\frac{1}{\varepsilon_1}$, on which approximating the regularised system, by averaging, is of practical validity. The size of the limit cycle along $x_1 = 0$ relates to the closed-loop system accuracy. The next Proposition shows that this size depends on the perturbation characteristics.
	\end{remark}
	\begin{proposition}\label{prop:bound_period}
		After the trajectories of the closed-loop system converge to the limit cycle, the bound on the state $x_1$ varies proportionally to the perturbation bound $L$ and to the square of the perturbation period $T$.
	\end{proposition}
	\begin{figure}[t]
		\begin{center}
			\includegraphics[width = 0.425\textwidth]{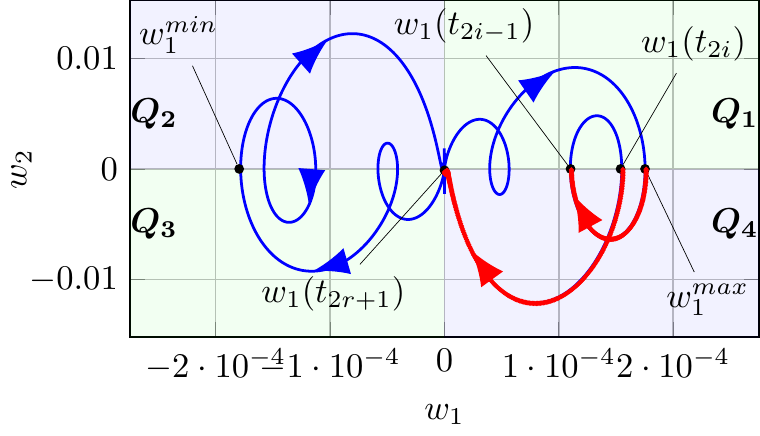}
			\vspace{-9pt}
			\caption{One full period of the limit cycle to which the trajectories of the closed-loop system in \eqref{eq:closed_loop_system} converge.}\label{fig:onePeriod}
		\end{center}
	\end{figure}
	\begin{pf}
		Denote the four quadrants of the phase space with $Q_j, \; j = 1,\dotsc, 4$. Introducing the new coordinates $w_1 \triangleq x_1$ and $w_2 \triangleq \dot{x}_1$ allows for re-writing \eqref{eq:closed_loop_system} as:
		\begin{align}
			\dot{w}_1 &=  w_2 \label{eq:w_1_dot}\\
			\dot{w}_2 &=  -\frac{1}{2}k_1\vert w_1 \vert^{-\frac{1}{2}}w_2 - k_2\text{sgn}(w_1) + q(t) \label{eq:w_2_dot}
		\end{align}
		Consider one period of the limit cycle as shown in Figure \ref{fig:onePeriod} restricted in $Q_1,Q_4$. For $t \geq t_0$ assume that the trajectories $\boldsymbol{w}(t)$ intersect with semi-axis $w_1 \geq 0$ at $2r + 1$ points, $r\in\mathbb{N}$ starting from $w_1^{max} \triangleq w_1(t_0)$, which is the maximum value of $w_1(t)$. Since the trajectories cannot cross from $Q_1$ to $Q_2$ (due to increasing $w_1$), each trajectory segment that lies in $Q_4$ starting at an intersection point will have to either cross the semi-axis $w_1 \geq 0$ twice (one while crossing to $Q_1$ and one right after while crossing to $Q_4$) or cross the vertical axis towards $Q_3$. In both cases, there will always be an odd number of intersections with the semi-axis $w_1 \geq 0$. Let these intersections occur at time instances $t_{2i}$ (from $Q_1$ to $Q_4$) and $t_{2i+1}$ (from $Q_4$ to $Q_1$), $i\in\mathcal{I} \triangleq \{0,\dotsc,r\}$ with $\boldsymbol{w}(t)$ crossing from $Q_4$ to $Q_3$ at $t = t_{2r+1}$. In each time interval $[t_{2i},t_{2i+1}]$, where $w_1(t) > 0, w_2(t) \leq 0$ holds (red lines):
		\begin{align}
			\dot{w}_2(t) &= -\frac{1}{2 \sqrt{w_1(t)}}k_1w_2(t) - k_2 + q(t) \geq -(k_2 + L) \Rightarrow \nonumber\\
			w_2(t) &\geq -(k_2 + L)(t - t_{2i}), \; \forall t\in(t_{2i},t_{2i+1}] \label{eq:w_2_t_3_bound}
		\end{align}
		since $w_2(t_{2i}) = 0$. The previous inequality leads to
		\begin{align*}
			\int_{t_{2i}}^{t_{2i+1}}w_2(t)dt &> -\frac{1}{2}(k_2 + L)(t_{2i+1} - t_{2i})^2, \; i\in\mathcal{I}.
		\end{align*}
		Hence, since $w_1(t_{2r+1}) = 0$ and $w_2(t) \geq 0, \; \forall t\in[t_{2i-1},t_{2i}], \; i\in\mathcal{I}-\{0\}$, it follows that
		\begin{align*}
			&-w_1^{max} = \int_{t_0}^{t_{2r+1}}w_2(t)dt = \sum\limits_{i=0}^{r}\int_{t_{2i}}^{t_{2i+1}}w_2(t)dt\\
			&+ \sum\limits_{i=1}^{r}\int_{t_{2i-1}}^{t_{2i}}w_2(t)dt \geq -\frac{1}{2}(k_2 + L)\sum\limits_{i=0}^{r}(t_{2i+1} - t_{2i})^2\\
			&> -\frac{1}{2}(k_2 + L)\left( \sum\limits_{i=0}^{r}(t_{2i+1} - t_{2i}) \right)^2 > -\frac{1}{8}(k_2 + L)T^2
		\end{align*}
		given that $t_{2i+1} > t_{2i}, \; \forall i\in\mathcal{I}$ and $\sum\limits_{i=0}^{r}(t_{2i+1} - t_{2i}) < t_{2r+1} - t_0 < \frac{T}{2}$. This leads to $w_1^{max} < \frac{1}{8}(k_2 + L)T^2$. Due to the homogeneity of the \gls{STSMC} closed-loop system \cite{seeber2018necessary}, by following the same reasoning in $Q_2$ one obtains $-w_1^{min} < \frac{1}{8}(k_2 + L)T^2$, where $w_1^{min} < 0$ is the minimum value that $x_1$ assumes. Finally, combining the two inequalities leads to
		\begin{equation}\label{eq:w_1_bound_1}
			\max\limits_{x(t)\in\gamma_{\varepsilon}(t)} \vert x_1(t) \vert = \max \left( w_1^{max}, -w_1^{min} \right) < \frac{1}{8}(k_2 + L)T^2 \; ,
		\end{equation}
		which completes the proof. \qed
	\end{pf}
	\begin{remark}\normalfont
		Proposition \ref{prop:bound_period} implies that the perturbation effect on the system vanishes as $T\rightarrow 0$. This can be interpreted as the perturbation not having ``enough time" to affect the system compared to the system's dynamics time scale.
	\end{remark}

\section{Tuning guidelines}	\label{sec:tuningRules}
	A geometric approach is followed in this section to derive bounds for $x_1 = w_1$. These bounds are expressed as functions of the controller gains $k_1,k_2$, the perturbation bound $L$ and its period $T$. The importance of this derivation is twofold: The dependence of the bound on the gains provides some guidelines for the tuning of the \gls{STSMC}. Conversely, the perturbation parameters can be used for predicting the closed-loop system performance given a specific set of controller gains.

\begin{figure}[t]
	\begin{center}
		\includegraphics[width = 0.45\textwidth]{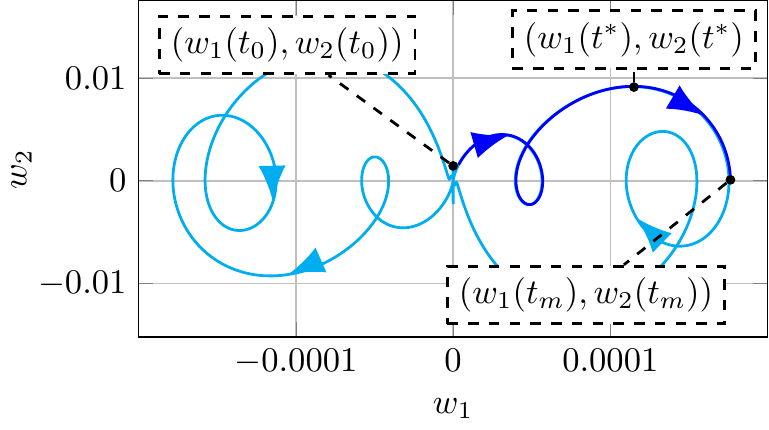}
		\caption{One full period of the limit cycle to which the trajectories of the closed-loop system in \eqref{eq:closed_loop_system} converge.}\label{fig:onePeriodFull}
	\end{center}
\end{figure}

Consider the limit cycle $\gamma_{\varepsilon}(t)$ to which the trajectories of the closed loop system \eqref{eq:w_1_dot}-\eqref{eq:w_2_dot} converge, restricted to the time interval $[t_0,t_m]$ as shown in Figure \ref{fig:onePeriodFull}, where $t_0 = 0$ is the instant when the trajectories cross from $Q_2$ to $Q_1$ and $w_1(t_m) = w_1^{max}$. Let $t^*\in [t_0,t_m)$ be the time instant where $w_2$ assumes its maximum value, i.e. $w_2(t^*) \geq w_2(t), \; \forall t\in [t_0,t_m]$. Integrating Equation \eqref{eq:w_2_dot} over the interval $[t_0,t_m]$, where $w_1(0) = 0, \; w_2(0) > 0$, $w_1(t_m) = w_1^{max}, \; w_2(t_m) = 0$ and $\dot{w}_2(t^*) = 0$ leads to
\begin{align}
	&\int_{0}^{t_m}\dot{w}_2(t)dt = \int_{0}^{t_m}(q(t) - k_2)dt - \int_{0}^{t_m} k_1\frac{\dot{w}_1(t)}{2\sqrt{w_1(t)}}dt \nonumber\\
	&\Rightarrow -w_2(0) \leq -k_1\sqrt{w_1^{max}} + (L - k_2)t_m \Rightarrow \nonumber\\
	&\sqrt{w_1^{max}} \leq \frac{w_2(0) + (L - k_2)t_m}{k_1} \; . \label{eq:full_int}
\end{align}
Moreover, evaluating Equation \eqref{eq:w_2_dot} at $t = t^*$ gives
\begin{align}
	\frac{k_1 w_2(t^*)}{2\sqrt{w_1(t^*)}} &= q(t^*) - k_2 \Leftrightarrow w_2(t^*) = 2\frac{q(t^*) - k_2}{k_1}\sqrt{w_1(t^*)} \; . \label{eq:max_w}	
\end{align}
Since $w_2(t^*) \geq w_2(t), \; \forall t\in [t_0,t_m]$, Equation \eqref{eq:max_w} yields
\begin{align*}
	w_2(0) &\leq w_2(t^*) \leq \frac{2(L - k_2)}{k_1}\sqrt{w_1(t^*)} \leq \frac{2(L - k_2)}{k_1}\sqrt{w_1^{max}}% \label{eq:ineq_1}
\end{align*}
which combined with \eqref{eq:full_int} leads to
\begin{align}
	&\sqrt{w_1^{max}} \leq \frac{\frac{2(L - k_2)}{k_1}\sqrt{w_1^{max}} + (L - k_2)t_m}{k_1} \Rightarrow \nonumber\\
	&\left[ k_1 - \frac{2(L - k_2)}{k_1} \right]\sqrt{w_1^{max}} \leq k_1(L - k_2)nT, \; 0 < n \leq \frac{1}{2} \; , \label{eq:ineq_2}
\end{align}
since $0 < t_m \leq \frac{T}{2}$. Finally, selecting
\begin{equation}\label{eq:condition_k_1}
	k_1 > \sqrt{2(L - k_2)}
\end{equation}
facilitates estimating a bound for the width $w_1^{max}$ of the limit cycle along the $x_1$-axis through \eqref{eq:ineq_2}. Specifically,
\begin{equation}\label{eq:w_1_bound}
	w_1^{max} \leq \frac{(L - k_2)^2 n^2T^2}{\left[ k_1 - \frac{2(L - k_2)}{k_1} \right]^2} \triangleq W_1(k_1,k_2) \; .
\end{equation}
Carrying out the same calculations but in $Q_3$ leads to the same inequality. The bound obtained in \eqref{eq:w_1_bound} is a function of the \gls{STSMC} gains $k_1,k_2$ as well as of the perturbation bound $L$, its period $T$ and the parameter $n$ that expresses the time $t_m$ as a fraction of $T$. Although this bound is conservative, especially if $n = \frac{1}{2}$, it can provide a useful insight regarding the tuning of the closed loop. It can be easily seen from \eqref{eq:w_1_bound} that $\lim\limits_{k_1\rightarrow \infty}W(k_1,k_2) = 0$, which means that provided that $k_1,k_2$ satisfy conditions \eqref{eq:convergence_condition_k2}, \eqref{eq:convergence_condition_k1} and \eqref{eq:condition_k_1}, the gains can be gradually increased in order to achieve accuracy better than $W_1(k_1,k_2)$. Given a specific error bound $\eta > 0$, it is possible to employ numerical optimisation methods for obtaining $k_1^*, \; k_2^*$ such that $\vert W(k_1^*,k_2^*) - \eta \vert \leq \epsilon$, where $\epsilon > 0$ is some numerical tolerance.
\begin{remark}\normalfont \label{rem:chatter_bound}
	The level of chatter in the controlled variable $w_1$ and, by extension, in the control signal relates to the maximum rate of change of $w_1$, i.e. the bound on $w_2$. By following a similar argument as in inequality \eqref{eq:w_2_t_3_bound} but for the first quadrant in Figure \ref{fig:onePeriodFull}, it can be shown that $\vert w_2(t) \vert \leq \vert w_2(t^*) \vert \leq (k_2 + L)\frac{T}{2}$. This implies that if $k_2$ cannot be selected larger than $L$, increasing its value introduces a trade-off between smaller error and larger amount of chatter.
\end{remark}
\begin{remark}\normalfont
	It is important to note that the approximation of the bound holds only if $k_1 > \sqrt{2(L - k_2)}$, which is another tuning decision. If this condition on $k_1$ is disregarded, $k_1$ can be selected to satisfy \eqref{eq:convergence_condition_k1} and still achieve acceptable bounds on $x_1$. However, inequality \eqref{eq:w_1_bound} can no longer be used to estimate these bounds.
\end{remark}

\section{Simulation results} \label{sec:simulations}
	The tracking problem for the angular velocity $\omega$ of a motor that is perturbed by torque ripples is considered for illustrating the theoretical findings in simulation. Considering a constant angular velocity set point $\omega_r$, then the dynamics of the tracking error $e \triangleq \omega - \omega_r$ is given by
\begin{align*}
	\dot{e} &= \frac{1}{J}\left[ u - T_F(\omega) + d(t) \right]\\
	d(t) &= L_1\cos(\omega_r t) + L_2\cos(\omega_r 3 t), \; L_1 = 3 L_2 = -\frac{LJ}{2\omega_r},
	\end{align*}
where $ L > 0$, $T_F$ is the known friction torque and $d(t)$ is the periodic perturbation due to torque ripples (e.g. cogging and parasitic torques). Applying the control law
$$
	u = J\left[ -k^{\prime}_1\vert e \vert^{\frac{1}{2}}\phi_{\delta}(e,\delta) - k^{\prime}_2\int_0^t \phi_{\delta}(e(\tau),\delta)d\tau + T_F(\omega) \right]
$$
brings the closed-loop vector field in the form \eqref{eq:regularised_vector_field}, with $x_1 \triangleq e, k_1 \triangleq \frac{k^{\prime}_1}{J},k_2 \triangleq\frac{k^{\prime}_2}{J}$, $k^{\prime}_1, k^{\prime}_2 > 0$ and $q(t) \triangleq \frac{\dot{d}}{J} =  \frac{L}{2}\left[ \sin(\omega_r t) + \sin(\omega_r 3 t) \right]$, which is $T$-periodic with $T = \frac{2\pi}{\omega_r}$ and $\vert q(t) \vert \leq L$.
\begin{figure}[t]
	\begin{center}
		\includegraphics[width = 0.475\textwidth]{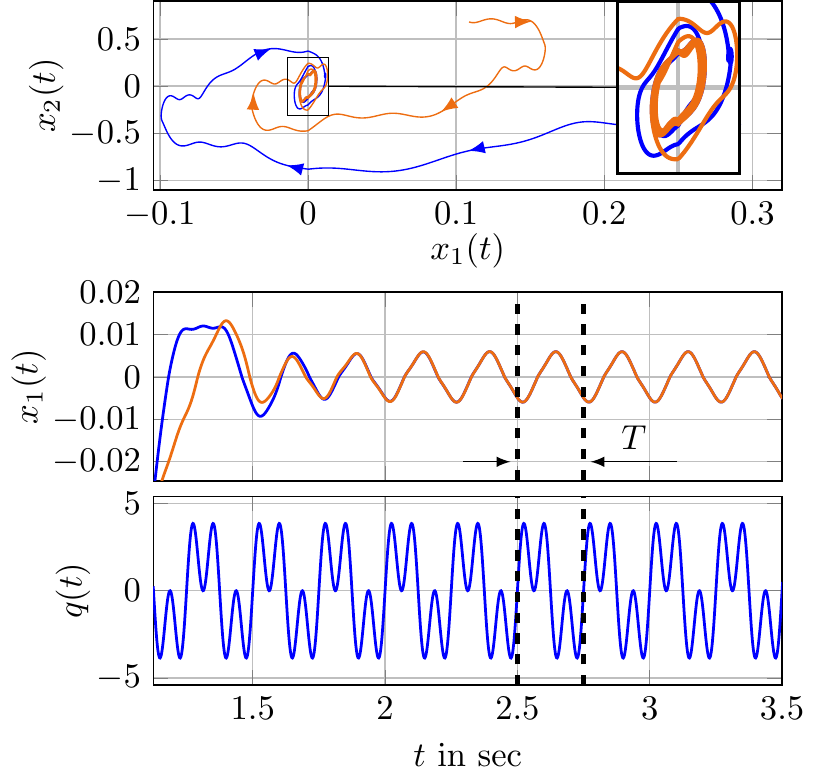}
		\vspace{-10pt}
		\caption{(Top) Starting from different initial conditions, the trajectories converge to a stable limit cycle of period $T = 0.25\si{\second}$. (Middle) Time response of $x_1(t)$ for two different initial conditions. (Bottom) Periodic perturbation rate $q(t)$.}\label{fig:full_convergence}
	\end{center}
\end{figure}

\begin{figure}[t]
	\begin{center}
		\includegraphics[width = 0.46\textwidth]{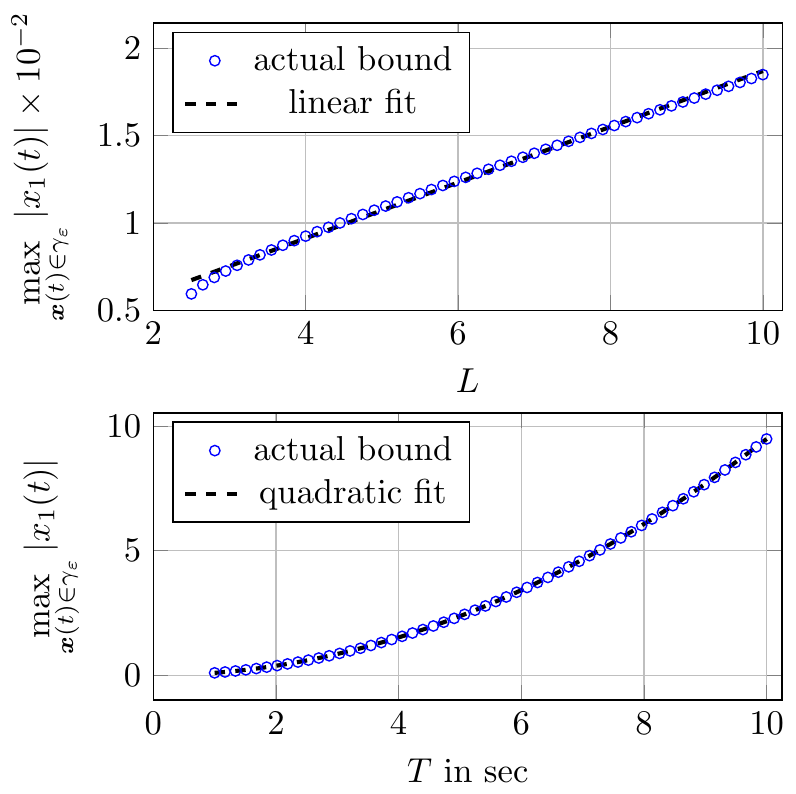}
		\caption{Bound on $x_1(t)$ as a function of the perturbation bound $L$ (top), and the perturbation period $T$ (bottom).}\label{fig:w_1_L_T}
	\end{center}
\end{figure}

Figure \ref{fig:full_convergence} shows the trajectories of the system's solutions with $L = 2.5$ and $T = 0.25 \si{\second}$ for different initial conditions. The gains were selected as $k_1 = 1$ and $k_2 = 2$, which satisfy conditions \eqref{eq:convergence_condition_k2} and \eqref{eq:convergence_condition_k1}. Proposition \ref{prop:bound_period} was also validated in simulation. The system was tested for different values of $L$ and $T$. As it can be seen in Figure \ref{fig:w_1_L_T}, the width of the limit cycle $w_1^{max}$ along the $x_1$-axis in the phase plot grows linearly with the bound $L$ of the perturbation and quadratically with the period $T$.
\begin{table}[t]\small
	\caption{Actual and estimated bound for the controlled variable $x_1$.}
	\label{tab:bounds_gains_tab}
	\centering
	%\begin{tabular}{cccccccc}\hline
%	$T$	& 	$L$ & $\bar{k}_1$ & $\bar{k}_2$ &  $k_1$ & $k_2$ & $\vert x_1 \vert$ & $W_1$\T\B \\
%	\specialrule{.2em}{.1em}{-1.2em}\\
%	$2$		& $2.5$ & $4.12$	& $2.75$ 	& $4.62$	& $0.54$ & $0.0066$ & $0.01$\T\B \\
%	$2$		& $25$ 	& $13.04$ 	& $27.5$	& $12.54$	& $12.9$ & $0.0055$ & $0.12$\T\B \\
%	$0.25$	& $2.5$ & $4.12$ 	& $2.75$ 	& $1.94$	& $0.88$ & $0.0002$ & $0.01$\T\B \\
%	$0.25$	& $25$ 	& $13.04$ 	& $27.5$	& $7.23$	& $5.85$ & $0.0035$ & $0.01$\T\B \\\hline
%\end{tabular}
\begin{tabular}{cccccccc}\hline
	$T$	& 	$L$ & $\bar{k}_1$ & $\bar{k}_2$ &  $k_1$ & $k_2$ & $\vert x_1 \vert$ & $W_1$\T\B \\
	\specialrule{.2em}{.1em}{-1.2em}\\
	$2$		& $2.5$ & $4.12$	& $2.75$ 	& $4.12$	& $0.43$ & $0.0099$ & $0.01$\T\B \\
	$2$		& $25$ 	& $13.04$ 	& $27.5$	& $12.54$	& $12.9$ & $0.0055$ & $0.12$\T\B \\
	$0.25$	& $2.5$ & $4.12$ 	& $2.75$ 	& $1.76$	& $1.08$ & $0.0002$ & $0.01$\T\B \\
	$0.25$	& $25$ 	& $13.04$ 	& $27.5$	& $6.14$	& $9.74$ & $0.0016$ & $0.01$\T\B \\\hline
\end{tabular}
\end{table}
Finally, the system was simulated under sinusoidal perturbation for two different values of magnitude $L\in\{ 2.5, 25 \}$ and two different periods $T\in\{ 2, 0.25 \}$. The objective of these tests was to compare the gains obtained by \eqref{eq:w_1_bound} for a given accuracy specification $\eta = 0.01$ with the ones required for finite-time stability. Moreover, the actual maximum deviation of $x_1$ from the origin was compared to the expected bound $W_1$. In all the tests, the gains $k_1,k_2$ were obtained by trying to solve the optimisation problem stated in Section \ref{sec:tuningRules}. Apart from the conditions \eqref{eq:convergence_condition_k2}, \eqref{eq:convergence_condition_k1} and \eqref{eq:condition_k_1}, additional saturation constraints were considered for the gains of the \gls{STSMC}. This was done to ensure that Assumption \ref{ass:contorller_gains} will not be violated. Table \ref{tab:bounds_gains_tab} shows the results of the simulations where $\bar{k}_1,\bar{k}_2$ denote the gains required for finite-time stability. The suggested values for $k_1$ are close to the ones of $\bar{k}_1$ for the slow perturbation but the ones for $k_2$ were chosen to be at most $50\%$ of those of $\bar{k}_2$ in all cases. The estimated bound is not always close to the prescribed accuracy $\eta$ due to the saturation constraints on the gains $k_1,k_2$. However, the actual bound for $x_1$ is up to 58 times smaller than the conservative estimation $W_1$.

\section{Conclusions} \label{sec:conclusions}
	The stability of under-tuned super-twisting sliding mode control loops under the effect of periodic perturbations were studied in this paper. Based on arguments from regularisation and averaging theory, it was shown that under milder gain conditions than the ones required for finite-time stability, the solutions of the closed-loop system, for sufficiently fast periodic perturbations, converge to a stable limit cycle. The width of the limit cycle in the phase plane linearly increases with the perturbation bound, while it quadratically decreases for smaller perturbation periods. Moreover, tuning guidelines were provided, based on a conservative estimation of the bound of the control variable. The theoretical findings were verified in simulation. Experimental validation will be pursued in future work, especially in the context of motion control systems such as machine tool drive trains.

%\begin{ack}                               % Place acknowledgements
%	Partially supported by the Roman Senate.  % here.
%\end{ack}

\bibliographystyle{ieeetr}        % Include this if you use bibtex (plain for alphabetical)
\bibliography{mybibl}           % and a bib file to produce the 
                                 % bibliography (preferred). The
                                 % correct style is generated by
                                 % Elsevier at the time of printing.

\end{document}